\documentclass[12pt]{article}
\usepackage{enumerate}
\usepackage{natbib}
\usepackage{url} % not crucial - just used below for the URL 
\usepackage{xr} 
\usepackage{comment}
\usepackage{caption}
\usepackage{amsmath,amssymb,amsthm,bm}
\usepackage{xcolor} 
\usepackage{multirow}  
\usepackage{booktabs}
\usepackage{makecell}
\definecolor{myblue}{RGB}{0, 0, 188}
\definecolor{mygreen}{RGB}{0, 188, 0}
\usepackage[colorlinks=false]{hyperref} % 
\usepackage{algorithm}
\usepackage{algpseudocode}

\usepackage{graphicx}
\usepackage{mathabx}
\usepackage{xr} 
\newcommand{\blind}{1}

\newtheorem{theorem}{Theorem}[section]
\newtheorem{lemma}[theorem]{Lemma}
\newtheorem{proposition}[theorem]{Proposition} 
\newtheorem{remark}{Remark}[section]

\newtheorem{assumption}{Assumption}[section]

\begin{document}

\def\spacingset#1{\renewcommand{\baselinestretch}%
{#1}\small\normalsize} \spacingset{1}

%%%%%%%%%%%%%%%%%%%%%%%%%%%%%%%%%%%%%%%%%%%%%%%%%%%%%%%%%%%%%%%%%%%%%%%%%%%%%%

\if1\blind
{
  \title{\bf  Causal Network Discovery from Interventional Count Data with Latent Linear DAGs}
  \author{Yijiao Zhang and Hongzhe Li 
  \thanks{Corresponding author. Yijiao Zhang is a postdoctoral fellow and Hongzhe Li is Perelman Professor in the Department of Biostatistis, Epidemiology and Informatics at the University of Pennsylvania. 
    }\\
  Department of Biostatistics, Epidemiology and Informatics\\
  University of Pennsylvania}
  
      \date{}
  \maketitle
} \fi

\if0\blind
{
  \bigskip
  \bigskip
  \bigskip
\begin{center}
    \setlength{\baselineskip}{1.5\baselineskip}
    {\LARGE\bf  Causal Network Discovery from Interventional Count Data with Latent Linear DAGs}
\end{center}
  \medskip
} \fi
\bigskip
\begin{abstract}
    The increasing availability of interventional data offers new opportunities for causal discovery, with gene perturbation studies providing a prominent example. Such data are typically count-valued and subject to substantial measurement error arising from technical variability and latent state heterogeneity. Motivated by these challenges, we study identification and estimation in latent linear structural causal models for interventional count data. We propose a latent linear Gaussian directed acyclic graph (DAG) model with Poisson measurement error that explicitly separates the latent causal structure from the observed counts. { Under one-sparse mechanism shifts, allowing the structural coefficients and intercept of the targeted gene to change, we establish population-level identifiability of the full latent causal DAG when every node is covered by at least one non-vanishing targeted intervention.} Building on these identification results, we develop an estimation procedure based on sparse inverse matrix estimation and provide theoretical guarantees on estimation error and finite-sample causal discovery. Simulation studies and applications to  Perturb-seq data demonstrate the practical effectiveness of the proposed method.
\end{abstract}
 
\noindent%
	\vfill

\spacingset{1.9} % DON'T change the spacing!

\newpage
\section{Introduction}
Understanding causal relationships among variables is a central problem across many scientific disciplines, including biology, medicine, and economics. Directed acyclic graphs (DAGs) provide a widely used framework for representing causal systems. However, observational data alone are generally insufficient to identify the underlying causal DAG. In general, they determine only a Markov equivalence class (MEC), within which the directions of edges that are crucial for causal interpretation are not uniquely identifiable.

The increasing availability of interventional data offers new opportunities for identifying causal graphs. A prominent example arises in gene regulatory network inference, where recent CRISPR-based perturbation technologies \citep{shalem2014genome} have enabled Perturb-seq experiments that combine pooled gene perturbations with single-cell RNA sequencing, yielding high-throughput interventional expression data with known targets at single-cell resolution \citep{replogle2022mapping}. This experimental design provides rich interventional information for studying causal regulatory relationships.

Despite these advantages, analyzing Perturb-seq data poses several substantial challenges. First, gene expression measurements in single-cell perturbation experiments are inherently count-valued, reflecting sequencing read counts rather than continuous expression levels. Most existing interventional causal discovery methods rely on Gaussian or continuous approximations and are therefore not well suited for such data. Additionally, observed counts exhibit substantial technical noise, arising from library size effects and experimental batch effects, which induces non-negligible measurement error relative to the underlying latent expression. Ignoring this measurement structure can lead to biased inference and misleading causal conclusions \citep{zhang2018causal,sarkar2021separating}, yet it is rarely addressed in existing interventional methods.

Moreover, Perturb-seq data may exhibit unmeasured confounding arising from latent cellular states and shared regulatory programs, which can simultaneously influence multiple genes. Such latent confounding induces spurious dependencies among genes and violates the causal sufficiency or independent noise assumptions commonly adopted by existing interventional causal discovery methods, potentially leading to biased DAG structure learning.

Furthermore, in Perturb-seq experiments, the number of cells available for each individual perturbation is often limited.   {While many interventional causal discovery methods establish population-level identifiability, their finite-sample performance may rely on stable estimation of quantities such as score functions.} Such estimation can become particularly challenging in high-dimensional settings with small per-intervention sample sizes, limiting the applicability of existing methods to realistic Perturb-seq data.

\paragraph{Our contribution}
Motivated by these considerations, we focus on the problem of identification and estimation of latent linear causal DAG models for interventional count data with measurement errors. Our main contributions are threefold.

Methodologically, we propose a latent linear Gaussian DAG model with a Poisson measurement layer, which decouples the causal structure among latent expression levels from the observed measurement process. { Under one-sparse mechanism-shift interventions, including mean-shift and coefficient-shift soft interventions as well as hard interventions}, we establish population-level identifiability of the latent causal DAG while allowing for the presence of latent confounding. We further characterize the worst-case necessity of full intervention coverage for exact identification. The identification argument also extends to other coordinate-aligned measurement models under which the latent mean contrasts are identifiable. Our results do not rely on the causal faithfulness assumption, which can be restrictive in high dimensions \citep{uhler2013geometry}.

Computationally, building on these identifiability results, we develop a new estimation procedure for the latent causal coefficient matrix that integrates ideas from sparse inverse matrix estimation with explicit DAG constraints. The resulting optimization problem is efficiently solved using the alternating direction method of multipliers.

Theoretically, we derive non-asymptotic estimation error bounds for the proposed estimator of the causal coefficient matrix. Under a beta-min type condition on the interventional strength and signal strength on the true edges, we characterize the finite-sample DAG discovery rate and quantify the probability of exact DAG recovery.

\paragraph{Related Work} Below, we provide a non-exhaustive review of the most closely related work on causal discovery. For more comprehensive surveys, see \citep{squires2023causal}. Causal discovery from observational data has been extensively studied via \textit{constraint-based} and \textit{score-based} methods. \textit{Constraint-based} methods, such as the PC algorithm \citep{spirtes2000causation}, rely on conditional independence testing and typically rely on strong faithfulness assumptions. \textit{Score-based} methods formulate DAG learning as the optimization of a scoring function, ranging from greedy search \citep{chickering2002optimal} to continuous optimization \citep{zheng2018dags,bello2022dagma}. These approaches avoid faithfulness assumptions and instead require beta-min conditions for consistent recovery \citep{sara2013l0}. Purely observational data identify a Markov equivalence class, unless additional distributional assumptions are imposed, such as non-Gaussian errors \citep{Shimizu2006lingam} or equal error variances \citep{peters2014identifiability}.

Interventional data provide additional information for causal structure identification.
For known-target hard interventions, \citet{hauser2012characterization} characterized interventional Markov equivalence classes, \citet{wang2017permutation} developed a hybrid method combining structure optimization and conditional independence testing, and \citet{xue2025dotears} extended continuous score-based optimization to interventional settings.
{ A growing literature also studies soft interventions, particularly in causal representation learning (CRL), where intervention-induced distributional changes across environments are exploited to reduce latent-representation ambiguities and identify latent causal structure under suitable conditions
\citep{squires2023linear,buchholz2023learning, zhang2023identifiability,varici2025score}.}
Our framework is related to CRL, with the latent coordinates naturally anchored by the gene-wise correspondence between observed and latent variables.
The mean-shift and coefficient-shift interventions considered here can be viewed as structured single-target soft interventions within this broader class.

Several lines of work study causal discovery with latent confounding. One line considers acyclic directed mixed graphs, including constraint-based methods such as FCI and its variants \citep{spirtes2000causation,colombo2012learning}, score-based methods \citep{bhattacharya2021differentiable}, and approaches exploiting non-Gaussianity \citep{wang2023causal}. Another line performs deconfounding under structured error covariance assumptions \citep{frot2019robust,pal2025dag}. GrIVET \citep{chen2024grivet} uses intervention variables as instrumental variables (IVs) for local causal discovery with latent confounding. In contrast, our method leverages intervention-induced mean contrasts for global DAG estimation while allowing a non-diagonal structural error covariance, without explicitly estimating the latent confounding structure.

Most existing methods are not tailored to count-valued data. Several works study causal discovery for observational Poisson models \citep{park2015learning,park2019high} and zero-inflated Poisson models \citep{choi2023model}, but they do not account for latent confounding or measurement error.  A smaller body of work considers DAG learning in the presence of measurement error \citep{zhang2018causal,dai2022independence,saeed2020anchored}, which is largely constraint-based and restricted to observational settings.

%In particular, full DAG identifiability has been established for Poisson structural causal models \citep{park2019high}, exploiting the equidispersion property of the Poisson distribution. 

\paragraph{Organization} The remainder of the paper is organized as follows. We introduce the model, identifiability results, and estimation procedure in Sections \ref{sec:method} - \ref{subsec:alg}. Section \ref{sec:theory} develops theoretical guarantees, including estimation error bounds and finite-sample DAG identification results. Simulation studies are reported in Section \ref{sec:simu}, followed by real data applications in Section \ref{sec:appl}. Section \ref{sec:disc} concludes with a discussion.

\section{Statistical Model for Perturb-seq Single Cell RNA-seq Data}\label{sec:method}
We first introduce some otation. Let $[p]=\{1,\ldots,p\}$. For a matrix $B=(B_{jk})\in\mathbb{R}^{p\times p}$, define the max norm $\|B\|_{\max}=\max_{j,k}|B_{jk}|$, the induced matrix norms $\|B\|_{L_1}=\max_{k\in[p]}\sum_{j=1}^p|B_{jk}|$ and $\|B\|_{L_\infty}=\max_{j\in[p]}\sum_{k=1}^p|B_{jk}|$, the spectral norm $\|B\|_2=\sup_{\|x\|_2=1}\|Bx\|_2$, the elementwise $\ell_1$ norm $\|B\|_1=\sum_{j,k}|B_{jk}|$, and the Frobenius norm $\|B\|_F=(\sum_{j,k}B_{jk}^2)^{1/2}$. We denote by $\|B\|_0=|\{(j,k):B_{jk}\neq0\}|$ the number of nonzero entries in $B$.

We consider gene expression count data collected under multiple interventional environments generated by CRISPR perturbations.  Let $X^{(m)} = (X^{(m)}_1, \ldots, X^{(m)}_p)^\top \in \mathbb{R}^p$ denote the observed gene expression vector in environment $m \in \{0,1,\ldots,M\}$, where $m=0$ corresponds to the control (non-intervened) condition. 
Each interventional environment $m>0$ targets a single gene for perturbation, and we denote by 
$I^{(m)}=\ell_{m}\in [p]$ the index of the perturbed gene in environment $m$.

To separate the measurement process from the underlying expression process in single-cell RNA sequencing data \citep{sarkar2021separating}, we model the observed gene expression counts through an explicit measurement layer and a latent expression layer. Specifically, for each environment $m$, the observed count $X^{(m)}_j$ of gene $j$
is modeled as a Poisson random variable conditional on
the library size $L^{(m)}\in\mathbb{R}^{+}$, the cell-level covariates $C^{(m)} \in \mathbb{R}^d$, and the latent true expression level $Z^{(m)}_j$:
\begin{equation}\label{eq:obs-model}
X^{(m)}_j \mid L^{(m)}, C^{(m)}, Z^{(m)}_j 
\sim 
\operatorname{Poisson}\!\left(
L^{(m)}\exp\{s_j(C^{(m)})+Z^{(m)}_j\}
\right),
\quad j=1,\ldots,p.
\end{equation}
Here, $L^{(m)}$ accounts for sequencing depth, while $s_j(C^{(m)})\in\mathbb{R}$ captures the systematic effects of observable covariates such as batch effects that may effect the observed sequence counts for gene $j$. The function $s_j(\cdot)$ is shared across environments. After adjusting for technical effects and observable cell heterogeneity, $Z^{(m)}_j$
represents the latent expression level. { From a causal representation learning perspective, the gene-wise correspondence between $X_j$ and $Z_j$ provides a natural alignment of the latent coordinates and substantially reduces representation ambiguity.} We assume that $Z^{(m)}\perp (C^{(m)},L^{(m)})$ for all $m\in\{0,\ldots,p\}$, implying that the latent expression process is independent of the sequencing depth and possible batch effects.

To characterize the latent causal structure underlying gene expression, 
we consider a linear Gaussian structural causal model (SCM) on the latent expression vector 
$Z^{(m)}=(Z^{(m)}_1,\ldots,Z^{(m)}_p)^\top\in\mathbb{R}^p$ for each environment $m\in \{0,\ldots,M\}$. Specifically, we consider
{ 
\begin{equation}\label{eq:latent-model} Z^{(m)} \;=\; A^{(m)} Z^{(m)} \;+\; \eta^{(m)} \;+\; \varepsilon^{(m)}, \qquad \varepsilon^{(m)} \sim \mathcal{N}\!\big(0,\Sigma^{(m)}_e\big), \qquad (m=0,\ldots,M), \end{equation}}
where $A^{(m)}\in \mathbb{R}^{p\times p}$ is the structural coefficient matrix and $\eta^{(m)}\in\mathbb{R}^{p}$ is an environment-specific intercept vector. The exogenous noise $\varepsilon^{(m)}$ is modeled as Gaussian with environment-specific covariance $\Sigma^{(m)}_e$, which may be non-diagonal, allowing possible latent confounding. We write $A:=A^{(0)}$ for the structural coefficient matrix of primary interest. 
The sample size under the $m$-th environment is denoted by $n_m$.

The control coefficient matrix $A$ induces a directed graph
$\mathcal{G}=([p],\mathcal{E})$ on the vertex set $[p]=\{1,\ldots,p\}$, where
$\mathcal{E}=\{(i,j):A_{ji}\neq 0\}$.
In particular, a nonzero entry $A_{ji}$ corresponds to a directed edge $i\to j$ in $\mathcal{G}$.
We assume that $\mathcal{G}$ is acyclic, so that it forms a DAG.
Thus, $A$ represents the latent causal mechanism under the control environment. More generally, for any matrix $A\in\mathbb{R}^{p\times p}$, we denote by 
$\mathcal{G}(A)$ the directed graph induced by $A$.

We next characterize how an intervention modifies the latent structural mechanism. 
Let $e_j\in\mathbb{R}^{p}$ denote the $j$-th canonical basis vector.
For each interventional environment $m>0$ targeting gene $\ell_m$, we assume a 
\textit{one-sparse mechanism shift}, in the sense that only the node-specific structural mean mechanism of the targeted gene is allowed to change:
{ \begin{equation}\label{eq:sparse-mechanism-shift}
A^{(m)}
=
A^{(0)}
+
e_{\ell_m}\{\delta^{(m)}\}^{\top},
\qquad
\eta^{(m)}
=
\eta^{(0)}
+
\alpha_m e_{\ell_m},
\qquad
(m=1,\ldots,M).
\end{equation}
Here, $\delta^{(m)}\in\mathbb{R}^{p}$ characterizes the intervention-induced changes in the existing incoming structural coefficients of the targeted gene. We assume that $\delta_j^{(m)}\neq 0$ only when $A_{\ell_m j}\neq 0$, so that the intervention does not introduce new parents. Consequently, $\mathcal{G}(A^{(m)})$ is a subgraph of the control DAG and remains acyclic. }The scalar $\alpha_m\in\mathbb{R}$ denotes the corresponding intercept shift. 
Thus, the structural mean equations of all non-target genes remain invariant across environments.

{ 
\begin{remark}[Special cases of one-sparse mechanism shifts]
Equation~\eqref{eq:sparse-mechanism-shift} accommodates both soft and hard single-target interventions. For soft interventions, $\delta^{(m)}=0$ gives a pure mean shift \citep{li2026root}, with
$A^{(m)}=A$ and $\eta^{(m)}=\eta^{(0)}+\alpha_m e_{\ell_m}$, shifting the targeted gene without altering its incoming structural coefficients. In contrast, $\delta^{(m)}\neq0$ allows the structural coefficients of the targeted gene with respect to its existing parents to change \citep{buchholz2023learning}. These cases can represent gene activation or partial suppression. A hard intervention has $\{\delta^{(m)}\}^{\top}=-A_{\ell_m,\cdot}$, so that
$A^{(m)}_{\ell_m,\cdot}=0$ \citep{xue2025dotears}, removing all incoming causal effects and providing a natural representation of gene knockout.
\end{remark}
}

The acyclicity of $\mathcal{G}$ means there exists a permutation matrix $P$ (a topological order of $\mathcal{G}$) such that 
$P^\top A P$ is strictly lower triangular. In particular, $I-A$ is invertible and hence we define $B:=(I-A)^{-1}$. Let $\mu^{(m)}:=\mathbb{E}[Z^{(m)}]$. Under the one-sparse mechanism shift in \eqref{eq:sparse-mechanism-shift}, the latent mean difference between environment $m$ and the control satisfies
{ \begin{equation}\label{eq:Delta-mu}
\Delta\mu^{(m)}
\;:=\;
\mu^{(m)}-\mu^{(0)}
\;=\;
\gamma_m B e_{\ell_m},
\qquad
\gamma_m
:=
\alpha_m+\{\delta^{(m)}\}^{\top}\mu^{(m)}.
\end{equation}
Equation~\eqref{eq:Delta-mu} shows that the one-sparse mechanism shift at the targeted gene can propagate through the causal system and induce potentially dense mean changes in downstream genes.}

\section{Identification of the DAG Under Unmeasured Confounders}\label{subsec:identification}
Under linear Gaussian SCMs, observational data alone only identify the causal structure up to a Markov equivalence class. By contrast, the one-sparse mechanism shifts across interventional environments provide identifying information through cross-environment changes in the latent means. { Importantly, exploiting these mean contrasts does not require point identification of the cell-level latent expression vector $Z$, which is not identifiable from a single observed count vector under the stochastic observation model.} A key observation from \eqref{eq:Delta-mu} is that the latent mean contrast $\Delta\mu^{(m)}$ equals the $\ell_m$-th column of $B$ up to the scalar factor $\gamma_m$. Moreover, under acyclicity, $P^{\top}BP$ is unit lower triangular. This unit-diagonal constraint fixes the scale and sign of each column of $B$, so that the $\ell_m$-th column of $B$ is identifiable from $\Delta\mu^{(m)}$ whenever $\gamma_m\neq 0$.
Provided that for each node $\ell\in[p]$ there exists at least one environment $m$ with $\ell_m=\ell$ and $\gamma_m\neq 0$, stacking the recovered columns yields identifiability of $B$ and hence of the causal mechanism $A$. It therefore suffices to identify the cross-environment latent mean contrasts $\{\Delta\mu^{(m)}\}_{m=1}^{M}$ from the observed data.

\begin{assumption}
\label{assump:covariate-identification}
For each $j\in[p]$, the shared covariate effect $s_j(\cdot)$ is
identifiable from the collection of environments up to a gene-specific additive constant.
\end{assumption}

Assumption \ref{assump:covariate-identification} holds, for example, under sufficient overlap of the covariate distributions across environments. It may also hold without
such overlap under standard identifiability conditions for a parametric specification of $s_j(\cdot)$. 

Under the Poisson--lognormal observation model, moments of the observed counts admit an explicit mapping to moments of the latent expression variables
$Z^{(m)}$. { Specifically, letting
$S_j(C,L)=L\exp\{s_j(C)\}$ and
$Y_j^{(m)}=X_j^{(m)}/S_j(C^{(m)},L^{(m)})$, we have
\begin{equation}\label{eq:latent-mean}
\mu_j^{(m)}
=
\log\!\big(\mathbb{E}[Y_j^{(m)}]\big)
-\frac{1}{2}\Sigma_{jj}^{(m)},
\end{equation}
where
$\Sigma_{jj}^{(m)}
=
\log\!\left\{
\mathbb{E}\!\left[X_j^{(m)}(X_j^{(m)}-1)/S_j(C^{(m)},L^{(m)})^2\right]
/
\big(\mathbb{E}[Y_j^{(m)}]\big)^2
\right\}.$}
These relationships connect the latent mean $\mu^{(m)}$ to observable
moments of the covariate-adjusted counts.

Under Assumption~\ref{assump:covariate-identification}, the resulting
latent mean contrasts are uniquely identified from the observed
distribution. We then have the following identification result.

\begin{theorem}\label{thm:identification-1}
Under the model \eqref{eq:obs-model}--\eqref{eq:sparse-mechanism-shift}
and Assumption~\ref{assump:covariate-identification}, the causal
mechanism $A$ is identifiable by jointly leveraging observational and
interventional data, provided that each node is subject to at least
one non-vanishing one-sparse mechanism-shift intervention in the latent
expression model.
\end{theorem}
Theorem~\ref{thm:identification-1} establishes identifiability of the causal mechanism $A$ at the population level by jointly leveraging observational and interventional data. 
Here, a \emph{non-vanishing} mechanism-shift intervention refers to a perturbation with effective intervention strength $\gamma_m \neq 0$ in the latent expression model. While $\gamma_m\neq0$ suffices for population identification, finite-sample recovery requires $|\gamma_m|$ to be sufficiently large; see Section~\ref{sec:theory}.
{ 
The proof further shows that population-level identification does not require known intervention targets. Nevertheless, we focus on the known-target setting for estimation and finite-sample theory.

The measurement model is used to recover the latent mean contrasts from the observed counts, but does not by itself identify causal directions. The one-sparse interventions use these contrasts to recover the causal propagation encoded by $(I-A)^{-1}$. Once the latent mean contrasts are recovered, this identification argument no longer depends on the specific Poisson--lognormal observation model and can extend to other coordinate-aligned measurement models under which these contrasts are identifiable. Further connections to interventional CRL and possible extensions to less structured measurement models are discussed in Supplementary Section~A.2.

\begin{proposition}[\textbf{Worst-case necessity of full intervention coverage}]
\label{prop:full-coverage-necessity}
Under the model
\eqref{eq:obs-model}--\eqref{eq:sparse-mechanism-shift},
with positive-definite and possibly non-diagonal structural error
covariance matrices, full-node intervention coverage is necessary in
the worst case to guarantee exact identification of $A$ over the
considered model class. 
\end{proposition}
Importantly, this worst-case result does not imply that full intervention coverage is necessary for every particular DAG. Incomplete coverage can still provide informative constraints on the causal structure, as further discussed in Supplementary Section~B.5.}

\begin{table}[H]
\centering
\caption{Comparison of representative causal discovery methods.}
\label{tab:method_comparison}
{
\setlength{\tabcolsep}{3.2pt}
\renewcommand{\arraystretch}{0.66}
\resizebox{\textwidth}{!}{
\begin{tabular}{cccccccc}
\toprule
Method
& Data
& \makecell{Structural\\[-12pt]model}
& Intervention
& \makecell{Known\\[-12pt]targets}
& \makecell{Latent\\[-12pt]confounding}
& \makecell{Measurement\\[-12pt]model}
& \makecell{Identified object} \\
\midrule

PLNi
& Count
& LG
& \makecell{Mean shift\\[-10pt]Coef. shift\\[-10pt]Hard}
& Yes$^{*}$
& Yes
& Poisson
& Full latent DAG \\

Anchor-PLN-PC \citep{saeed2020anchored}
& Count
& LG
& None
& --
& No
& Poisson
& Latent MEC \\

MRS-PoissonSEM \citep{park2019high}
& Count
& Poisson GLM
& None
& --
& No
& None
& Full DAG \\

GrIVET \citep{chen2024grivet}
& Continuous
& LG
& Mean shift
& No
& Yes
& None
& Full DAG \\

GIES \citep{hauser2012characterization}
& Continuous
& LG
& Hard
& Yes
& No
& None
& i-MEC \\

IGSP \citep{wang2017permutation}
& Continuous
& General
& Hard
& Yes
& No
& None
& i-MEC \\

DOTEARS \citep{xue2025dotears}
& Continuous
& LG
& Hard
& Yes
& No
& None
& Full DAG \\

LiNGAM \citep{Shimizu2006lingam}
& Continuous
& LNG
& None
& --
& No
& None
& Full DAG \\

Contrastive-CRL \citep{buchholz2023learning}
& Continuous
& LG
& \makecell{Hard\\[-10pt]Soft}
& No
& No
& \makecell{Nonlinear\\[-10pt]mixing}
& \makecell{Full latent DAG$^{\dagger}$\\[-10pt]
Ancestral order} \\

\bottomrule
\end{tabular}
}}
\begin{flushleft}
\footnotesize
LG = linear Gaussian; LNG = linear non-Gaussian; GLM = generalized linear model; \\[-10pt]i-MEC = interventional Markov equivalence class. \\[-10pt]
$^{*}$ Targets are used for estimation and finite-sample theory, but not for population identification. \\[-10pt]
$^{\dagger}$ The causal graph is identified up to a permutation of the node labels.
\end{flushleft}
\end{table}

{ 
Table~\ref{tab:method_comparison} compares PLNi with representative existing methods along the key dimensions relevant to our identification setting.
} A key feature of our identification strategy is that it allows a non-diagonal structural error covariance. Rather than explicitly estimating the latent confounding structure, the proposed method exploits intervention-induced latent mean contrasts, which remain informative in the presence of correlated structural errors. This provides a global identification strategy for the directed causal structure without node-wise deconfounding or explicit covariance decomposition.

\section{Estimation of Causal Coefficient Matrix and DAG Recovery}\label{subsec:alg}
In this section, we describe the procedure for estimating the causal coefficient matrix $A$ and recovering the corresponding DAG, consisting of four main steps:
(i) estimating and removing measurement-layer effects,
(ii) mapping observed moments to latent means under the Poisson--lognormal model,
(iii) recovering the columns of $B=(I-A)^{-1}$ from cross-environment mean contrasts,
and (iv) estimating $A$ under sparsity and acyclicity constraints.
Algorithm~\ref{alg:PLN} summarizes the procedure, which we refer to as PLNi.

We begin by estimating the covariate effect $s_j(C)$ in the measurement layer. { Under the assumption $Z^{(m)}\perp(C^{(m)},L^{(m)})$, marginalizing over
$Z_j^{(m)}$ yields
\begin{equation}\label{eq:s_j}
\log \mathbb{E}[X^{(m)}_{ij}\mid C^{(m)}_i,L^{(m)}_i]
=
\log L^{(m)}_i
+
s_j(C_i^{(m)})
+
\xi_j^{(m)},
\end{equation}
where $\xi_j^{(m)}$ is an environment-specific marginal log-mean offset. Under Assumption~\ref{assump:covariate-identification}, $s_j(\cdot)$ is identifiable up to a gene-specific additive constant. For estimation, we therefore adopt an identifiable parameterization of $s_j(\cdot)$. This choice does not affect the subsequent cross-environment
latent mean contrasts. For example, one may use a linear specification $s_j(C^{(m)})=\beta_j^\top C^{(m)}$ estimated by Poisson quasi-maximum likelihood, or a flexible smooth regression such as spline-based generalized additive modeling. Let $\widehat s_j(C^{(m)})$ denote the resulting estimator and define
$\widehat S_{ij}^{(m)}
=
L_i^{(m)}\exp\{\widehat s_j(C_i^{(m)})\}$,
$\widehat Y_{ij}^{(m)}
=
X_{ij}^{(m)}/\widehat S_{ij}^{(m)}.$}
We then estimate $\widehat\mu^{(m)}$ using the empirical counterparts of \eqref{eq:latent-mean}, as specified in
\eqref{eq:est-latent-mean}.

For each interventional environment $m\in\{1,\ldots,p\}$, we compute latent mean contrasts $\Delta\widehat\mu^{(m)}=\widehat\mu^{(m)}-\widehat\mu^{(0)}$ and reconstruct the $\ell$-th column of $B$ by $\widehat{b}_{\ell} = \Delta\widehat\mu^{(m)}/\Delta\widehat\mu^{(m)}_{\ell}$ where $\ell=I^{(m)}$. Collecting these columns yields $\widehat{B}=[\widehat{b}_1,\ldots,\widehat{b}_p]\in\mathbb{R}^{p\times p}$. 

Finally, to recover the causal matrix $A$, we estimate a sparse weighted adjacency matrix $A$ under an acyclicity constraint. Given the estimate \(\widehat{B}\approx (I-A)^{-1}\), we define
\begin{equation}
\label{eq:main_opt}
\widehat A \in \arg\min_A\ \|A\|_{1}
\quad \text{s.t.}\quad 
h(A)=0,\quad 
\|(I-A)\widehat{B}-I\|_{\max}\le \lambda_s,
\end{equation}
where $h(A)=0$ enforces acyclicity, and the max-norm constraint
controls the approximation error of $(I-A)\widehat B$ through a
CLIME-type regularization \citep{cai2011clime}. Typical choices include the NOTEARS
characterization
$h(A)=\operatorname{tr}\{\exp(A\circ A)\}-p$
\citep{zheng2018dags} and the DAGMA characterization
$h(A)=-\log\det(uI-A\circ A)+p\log u$
on the domain $\varrho(A\circ A)<u$
\citep{bello2022dagma}, where $\varrho(\cdot)$ denotes the spectral radius. 
\begin{algorithm}[h]
	\caption{PLNi: Causal discovery from interventional Poisson count data}
	\label{alg:PLN} 
	\begin{algorithmic}
		\Require Observed data $\{(X^{(m)},L^{(m)},C^{(m)})\}_{m=0}^{M}$, intervention targets $\{I^{(m)}=\ell_m\}_{m=1}^{M}$, regularization level $\lambda_s$, threshold level $\tau>0$.\\
		
		\noindent 1. \underline{\textbf{Measurement-layer adjustment:}}
Estimate $\widehat s_j(C_i)$ from (\ref{eq:s_j}) and set
$\widehat S_{ij}^{(m)}
=
L_i^{(m)}\exp\{\widehat s_j(C_i^{(m)})\}$
and
$\widehat Y_{ij}^{(m)}
=
X_{ij}^{(m)}/\widehat S_{ij}^{(m)}$.\\
		
		\noindent 2. \underline{\textbf{Mapping observed moments to latent means:}}
Compute
\begin{equation}\label{eq:est-latent-mean}
\widehat\mu_j^{(m)}
=
\log\!\big(\widehat{\mathbb{E}}[\widehat Y_j^{(m)}]\big)
-\tfrac12\widehat\Sigma_{jj}^{(m)},
\quad
\widehat\Sigma_{jj}^{(m)}
=
\log\!\left\{
\frac{
\widehat{\mathbb{E}}\!\left[
X_j^{(m)}(X_j^{(m)}-1)/
(\widehat S_j^{(m)})^2
\right]
}{
\widehat{\mathbb{E}}[\widehat Y_j^{(m)}]^2
}
\right\}.
\end{equation}

        \noindent 3. \underline{\textbf{Column recovery of $B$:}} For the environment $m$ targeting ${\ell}=I^{(m)}$: reconstruct the $\ell$-th column of $B$ by $\widehat{b}_{\ell} = \Delta\widehat\mu^{(m)}/\Delta\widehat\mu^{(m)}_{\ell}$. Collect the columns into a matrix to obtain $\widehat{B}=[\widehat{b}_1,\ldots,\widehat{b}_p]\in\mathbb{R}^{p\times p}$.\\

        \noindent 4. \underline{\textbf{Sparse DAG estimation of $A$:}} Obtain $\widehat{A}$ by solving (\ref{eq:main_opt}).
		
		\Ensure
		Estimated causal matrix $\widehat A$ and recovered graph $\widehat{\mathcal{G}}=\mathcal{G}(\widehat{A}_{\tau})$.
	\end{algorithmic}
\end{algorithm}
To handle the nonsmooth max-norm constraint, we introduce
$Z=(I-A)\widehat B-I$ and rewrite \eqref{eq:main_opt}as
\begin{equation}
\label{eq:split_problem}
\min_{A,Z}\ \|A\|_1
\quad\text{s.t.}\quad
A\widehat B+Z=\widehat B-I,\qquad
\|Z\|_{\max}\le\lambda_s,\qquad
h(A)=0.
\end{equation}
We handle the acyclicity constraint $h(A)=0$ using an outer
augmented-Lagrangian procedure, with multiplier $\nu\ge0$ and
quadratic penalty parameter $\omega>0$. At each outer iteration, with $(\nu,\omega)$ fixed, the equality constraint in
\eqref{eq:split_problem} is handled using the alternating direction
method of multipliers (ADMM). Let
$\mathcal C_{\lambda_s}
=
\{Z\in\mathbb R^{p\times p}:\|Z\|_{\max}\le\lambda_s\}$,
and let $\iota_{\mathcal C_{\lambda_s}}$ denote its indicator
function. Introducing a dual variable $V\in\mathbb R^{p\times p}$,
the augmented Lagrangian is
$$\mathcal L_{\rho}(A,Z,V)
=
\|A\|_1
+
\iota_{\mathcal C_{\lambda_s}}(Z)
+
\nu h(A)
+
\frac{\omega}{2}h(A)^2
+
\frac{\rho}{2}
\left\|
Z-\{(I-A)\widehat B-I\}
+\frac{V}{\rho}
\right\|_F^2
-
\frac{1}{2\rho}\|V\|_F^2,$$
where $\rho>0$ is the ADMM penalty parameter. For fixed $(\nu,\omega)$, we update $Z$, $A$, and $V$ sequentially:
\begin{equation}
\label{eq:ADMM-updates}
\begin{aligned}
Z^{(t+1)}
&=
\Pi_{\mathcal C_{\lambda_s}}
\left(
\widehat B-A^{(t)}\widehat B-I-\frac{V^{(t)}}{\rho}
\right),\\
A^{(t+1)}
&\approx
\arg\min_A
\left\{
\frac{\rho}{2}\|A\widehat B-C^{(t+1)}\|_F^2
+
\nu h(A)
+
\frac{\omega}{2}h(A)^2
+
\|A\|_1
\right\},\\
V^{(t+1)}
&=
V^{(t)}
+
\rho
\left\{
A^{(t+1)}\widehat B
+
Z^{(t+1)}
-
(\widehat B-I)
\right\},
\end{aligned}
\end{equation}
where
$C^{(t+1)}
=
\widehat B-I-Z^{(t+1)}-V^{(t)}/\rho$.
The $A$-subproblem is solved using a backtracking proximal-gradient method. The inner iterations terminate when the primal, dual, and $A$-optimality residuals satisfy their prescribed tolerances.

At the outer level, let $A^{[k]}$ denote the $k$th accepted iterate.
If the stopping criteria are not yet met, a new candidate is accepted
only if its acyclicity measure contracts by a prescribed factor
$c_h\in(0,1)$ relative to $h(A^{[k]})$; otherwise, $\omega$ is increased
and the inner optimization is repeated. Once accepted as
$A^{[k+1]}$, the multiplier is updated as
$\nu\leftarrow\nu+\omega h(A^{[k+1]})$.
The outer procedure terminates when the acyclicity violation and the relative change between successive accepted iterates satisfy their prescribed tolerances. Further details are provided in Supplementary Section~A.1.   {We denote the numerical solution returned at termination by $\widehat A^{\mathrm{alg}}$}. For $\tau>0$, define its entrywise hard-thresholded version by $(\widehat A^{\mathrm{alg}}_\tau)_{ij} = \widehat a_{ij}^{\mathrm{alg}} \mathbf 1\{|\widehat a_{ij}^{\mathrm{alg}}|\ge\tau\}$. The estimated graph produced by the numerical procedure is then defined as $\widehat{\mathcal G}^{\mathrm{alg}} = \mathcal G(\widehat A^{\mathrm{alg}}_\tau)$.

Under standard regularity conditions, the backtracking proximal-gradient updates are descent steps and any accumulation point of the inner ADMM iterates with vanishing primal, dual, and $A$-optimality residuals satisfies the first-order conditions of the penalized inner problem. The accepted outer iterates geometrically reduce the acyclicity measure to the prescribed tolerance. Details are provided in Supplementary Section~D.1.

The regularization parameter $\lambda_s$ is selected using a pseudo-BIC (PBIC) criterion. For each candidate $\lambda_s$, let
$\widehat A^{\mathrm{alg}}(\lambda_s)$ denote the corresponding
numerical solution and define
\begin{equation}
\label{eq:PBIC}
\mathrm{PBIC}(\lambda_s)
=
\frac{
\|(I-\widehat A^{\mathrm{alg}}(\lambda_s))\widehat B-I\|_F^2
}{
\|\widehat B-I\|_F^2
}
+
C_w
\frac{
\|\widehat A^{\mathrm{alg}}(\lambda_s)\|_0
}{
p(p-1)/2
}.
\end{equation}
{ 
The first term measures lack of fit relative to the zero-graph
benchmark, while the second measures graph complexity relative to the maximum number $p(p-1)/2$ of edges in a DAG. Together, they provide a scale-free balance between model fit and graph sparsity, with $C_w>0$ controlling the relative weight placed on sparsity.
} In our implementation, we set $C_w=2$ and select $\lambda_s$ by
minimizing \eqref{eq:PBIC} over a prespecified grid.

\section{Theoretical Properties}\label{sec:theory}
In this section, we investigate the estimation error of the causal coefficient matrix and the graph recovery consistency for the proposed estimator in Section \ref{subsec:alg}. We first introduce additional notations.

Let $A^* \in\mathbb{R}^{p\times p}$ denote the true causal coefficient matrix under the control environment, $B^*:=(I-A^*)^{-1}$, and $\mathcal{G}^*=\mathcal{G}(A^*)$. Let $\mu^{(m)*}:=\mathbb{E}[Z^{(m)}]$ denote the true latent mean under environment $m$, and define $\Delta\mu^{(m)*}:=\mu^{(m)*}-\mu^{(0)*}$. For $\tau > 0$, define the entrywise hard-thresholded matrix $A_\tau$ by $(A_\tau)_{ij} := a_{ij}\,\mathbf{1}\big\{|a_{ij}|\ge \tau\big\},1\le i,j\le p$. Define $n_{\min}=\min_{m\in[p]\cup\{0\}} n_m$. We make the following assumptions.

\begin{assumption}[\textbf{Moments}]\label{assump:moment}
    Suppose there exist some constants $r,c, \delta,c_{\min}>0$ such that (i) $p\leq cn_{\min}^r$, (ii) $\mathbb{E}|Y_j^{(m)}|^{4+8r+\delta} \leq K$, and (iii) $\mathbb{E}[Y_j^{(m)}]\geq c_{\min}$ for $j\in[p]$ and $m\in[p]\cup\{0\}$.
\end{assumption}

Assumption~\ref{assump:moment} imposes moment conditions on the scaled observed Poisson counts to ensure well-behaved concentration of moment estimators. It is mild for Poisson-lognormal models, as Gaussian latent variables imply finite moments for the Poisson rates and the scaled counts; similar conditions are standard in related work \citep{cai2011clime,xiao2022estimating}. The lower bound $\mathbb{E}[Y_j^{(m)}]\ge c_{\min}$ excludes degenerate variables and ensures stable moment estimation.

{  
\begin{assumption}[\textbf{Nuisance Estimation}]
\label{assump:nuisance}
There exist a constant $c_s>0$, a deterministic sequence
$\epsilon_{s,n}>0$, and a sequence $\pi_{s,n}\to0$ such that
$\epsilon_{s,n}\le c_s\sqrt{\log p/n_{\min}}$ and
$P(\max_{\substack{0\le m\le p\\1\le i\le n_m\\1\le j\le p}}
\left|
\widehat s_j(C_i^{(m)})-s_j(C_i^{(m)})
\right|
\le \epsilon_{s,n})
\ge 1-\pi_{s,n}.$
\end{assumption}

Assumption~\ref{assump:nuisance} requires the uniform nuisance estimation error to be no larger than the stochastic rate $\sqrt{\log p/n_{\min}}$ of latent mean estimation. Since $s_j(\cdot)$ is shared across environments, it can be estimated by pooling all $N=\sum_{m=0}^{p}n_m$ observations. For example, under a low-dimensional linear specification $s_j(C)=\beta_j^\top C$, standard uniform estimation results give $\epsilon_{s,n}=O(\sqrt{d\log p/N})$ under suitable regularity conditions. Since $N\ge(p+1)n_{\min}$, this satisfies Assumption~\ref{assump:nuisance} when $d=O(p)$.
}

The following lemma establishes a uniform tail bound for the latent
mean estimation error across all variables and environments.

\begin{lemma}
\label{lem:est-hatmu}
Suppose Assumptions~\ref{assump:moment} and
\ref{assump:nuisance} hold. Then there exists a constant $C>0$ such
that, with probability at least
$1-C(n_{\min}^{-\delta/8}+p^{-t}+\pi_{s,n})$,
\begin{equation}
\max_{m,j}
\left|
\widehat\mu_j^{(m)}-\mu_j^{(m)*}
\right|
\le
c_\mu
\sqrt{\frac{\log p}{n_{\min}}},
\label{eq:hatmu-bound}
\end{equation}
where $c_\mu=c_1c_2+3c_s$ with
$c_2=
3/4
\left(
c_{\min}^{-1}
+
(4c_{\min}^2)^{-1}
\right)$.
\end{lemma}

\begin{assumption}[\textbf{Noise Nondegeneracy}]
\label{assump:noise}
Suppose that
$\lambda_{\min}(\Sigma_e^{(0)})\ge c_e$
for some constant $c_e>0$.
\end{assumption}

\begin{assumption}[\textbf{Intervention Strength}]
\label{assump:intervention-strength}
Let $\Delta_{\min}=\min_{m\in[p]}|\Delta\mu_{\ell_m}^{(m)*}|$.
Assume that
$\Delta_{\min}\ge 4c_\mu\sqrt{\log p/n_{\min}}$,
where $c_\mu$ is defined in Lemma~\ref{lem:est-hatmu}.
\end{assumption}

Assumption~\ref{assump:intervention-strength} specifies a minimal effective intervention strength for finite-sample recovery. Under the one-sparse mechanism-shift model,
$\Delta\mu_{\ell_m}^{(m)*}=\gamma_m^*$, so $\Delta_{\min}$ is the minimum effective first-moment intervention strength across environments.
The requirement
$\Delta_{\min}\gtrsim\sqrt{\log p/n_{\min}}$
matches the uniform stochastic error rate for estimating the latent means and therefore represents the natural finite-sample separation scale. Moreover, the condition can be empirically assessed using the estimated latent mean shifts across environments. 

The following theorem provides finite-sample error bounds for estimating the latent causal coefficient matrix $A^*$ under interventional Poisson count data. 
%The entrywise bound is the key rate for edge-wise accuracy and support recovery, while the Frobenius bound further quantifies global estimation error under weak row-sparsity $s_{\mathrm{in}}(p)$ for $q\in[0,1)$.The appearance of $\Delta_{\min}$ in the denominator reflects the impact of effective intervention strength on estimating the coefficient matrix $A^*$, as weaker intervention-induced mean contrasts reduce the separability between control and interventional data and necessarily inflate the estimation error. The dependence on $s_0(p)$ captures the effect of graph density, with larger $s_0(p)$ corresponding to denser underlying graphs and leading to increased estimation error in the spectral and Frobenius norms.

\begin{theorem}[\textbf{Estimation}] \label{thm:estimation} Assume that $\max\{\|I-A^*\|_{L_1},\|I-A^*\|_{L_\infty}\}\le R$ for some $R>0$. For each $j\in[p]$, let $\operatorname{An}^*(j)$ denote the set of ancestors of node $j$ in $\mathcal G^*$, and define $R_{\mathrm{anc}} = \max_j \sum_{k\in\operatorname{An}^*(j)\cup\{j\}} \|\boldsymbol a_k^*\|_1$. Under Assumptions~\ref{assump:moment}- \ref{assump:intervention-strength}, let $c_B= 4c_\mu\{1+ [\log(K^{2/(4+8r+\delta)}/c_{\min}^2)/c_e]^{1/2}\}$ and take $\lambda_s= c_B R\Delta_{\min}^{-1} \sqrt{\log p/n_{\min}}$. Then there exists a constant $C>0$ such that, with probability at least $1-C(n_{\min}^{-\delta/8}+p^{-t}+\pi_{s,n})$, the following statements hold: \[ \|\widehat A-A^*\|_{\max} \le \frac{ c_B R(R+1+R_{\mathrm{anc}}) }{ \Delta_{\min} } \sqrt{\frac{\log p}{n_{\min}}}. \] If, in addition, $A^*$ is weakly sparse in the sense that $\max_j\sum_{k=1}^p|a_{jk}^*|^q\le s_{\mathrm{in}}(p)$ for some $0\le q<1$, then \[ \frac{\|\widehat A-A^*\|_F^2}{p} \le 4s_{\mathrm{in}}(p) \left\{ \frac{ c_B R(R+1+R_{\mathrm{anc}}) }{ \Delta_{\min} } \sqrt{\frac{\log p}{n_{\min}}} \right\}^{2-q}. \] 
\end{theorem}
Theorem~\ref{thm:estimation} separates the effects of intervention strength, structural complexity, and sparsity on estimation accuracy. The entrywise bound characterizes edge-wise estimation accuracy, while the Frobenius bound further quantifies global estimation error under weak row sparsity. The factor $\Delta_{\min}^{-1}$ reflects the impact of effective intervention strength on estimating $A^*$, as weaker intervention-induced mean contrasts reduce the separability between control and interventional data and enlarge the estimation error bound. The quantity $R$ controls the overall coefficient magnitude, whereas $R_{\mathrm{anc}}$ captures the cumulative coefficient strength over ancestral sets and remains controlled when ancestral neighborhoods are small and their coefficient sums are bounded. Finally, $s_{\mathrm{in}}(p)$ determines how the entrywise error accumulates into the Frobenius error under weak sparsity.

To characterize the reconstructability of the underlying DAG, define
\[
\zeta
=
\inf_{\substack{
A:\,\mathcal G(A_\tau)\neq\mathcal G^*,\,
h(A)=0,\,
\|A\|_1\le\|A^*\|_1}}
\|(I-A)B^*-I\|_{\max}.
\]
The quantity $\zeta$ measures the population separation between the true DAG and acyclic competitors with an incorrect thresholded graph. To make this quantity more interpretable, let $a_{\min}:=\min_{(j,k):a_{jk}^*\neq0}|a_{jk}^*|$ denote the minimum nonzero edge strength and take $\tau=a_{\min}/2$.   {Then $\zeta\ge a_{\min}/\{2\|I-A^*\|_{L_1}\}\ge a_{\min}/(2R)$. Thus, a larger minimum edge strength increases the guaranteed separation, whereas poorer inverse stability of $B^*$, as measured by $\|(B^*)^{-1}\|_{L_1}$, weakens it. Graph density may also affect this stability through larger weighted coefficient sums.}

\begin{assumption}[\textbf{DAG Reconstructability}]
\label{assump:C_min}
Assume that
\[
\zeta
\ge
\frac{c_B}{\Delta_{\min}}
\left\{
R+
\frac{c_3(1+R_{\mathrm{anc}})}{c_\mu}
\right\}
\sqrt{\frac{\log p}{n_{\min}}},
\]
for some sufficiently large constant $c_3>0$.
\end{assumption}

Under the choice $\tau=a_{\min}/2$, the above lower bound on $\zeta$ implies that a sufficient condition for Assumption~\ref{assump:C_min} is \[ a_{\min} \ge \frac{2c_B R}{\Delta_{\min}} \left\{ R+ \frac{c_3(1+R_{\mathrm{anc}})}{c_\mu} \right\} \sqrt{\frac{\log p}{n_{\min}}}. \] This is a beta-min type condition similar to those commonly used in \citep{cai2011clime,sara2013l0,yuan2019constrained,park2019high}.

\begin{remark}
   We remark that the beta-min condition is not directly comparable to the strong faithfulness assumption; the former is a local signal strength condition commonly used in high-dimensional settings, whereas the latter is a global distributional assumption. See \citet{sara2013l0} for detailed discussion.
\end{remark}

\begin{theorem}[\textbf{Graph Recovery}]
\label{thm:graph-recovery}
Let $\lambda_s$ be chosen as in Theorem~\ref{thm:estimation}.
Under the conditions for the max-norm bound in
Theorem~\ref{thm:estimation}, suppose that $\zeta>\lambda_s$ and define
\[
\kappa
=
\frac{1}{8c_2}
\left\{
\frac{
c_\mu(\zeta-\lambda_s)\Delta_{\min}
}{
c_B(1+R_{\mathrm{anc}})
}
-
3c_s\sqrt{\frac{\log p}{n_{\min}}}
\right\}.
\]
For $\kappa>0$, there exists a constant $C>0$ such that
\begin{equation*}
\begin{aligned}
P(\widehat{\mathcal G}=\mathcal G^*)
\ge\;&
1-
8\exp\left[
-\frac{n_{\min}}{4}
\min\left\{
\frac{\kappa^2}{K},
3\kappa
\sqrt{\frac{(\log p)^3}{n_{\min}}}
\right\}
+2\log p
\right]-
C\left(
n_{\min}^{-\delta/8}
+
p^{-t}
+
\pi_{s,n}
\right),
\label{eq:graph-recovery-prob}
\end{aligned}
\end{equation*}
Further, under Assumption~\ref{assump:C_min}, $P(\widehat{\mathcal G}=\mathcal G^*)
\ge
1-C\left(
n_{\min}^{-\delta/8}
+
p^{-t}
+
\pi_{s,n}
\right),$
for some constant $C>0$, and consequently
$P(\widehat{\mathcal G}=\mathcal G^*)\to1$.
\end{theorem}
Theorem~\ref{thm:graph-recovery} establishes finite-sample recovery of the latent causal DAG via thresholding the estimated coefficient matrix. The convergence probability is governed by the degree of DAG reconstructability $\zeta$ and the intervention strength $\Delta_{\min}$ through $\kappa$: larger values of either quantity accelerate convergence of $P(\widehat{\mathcal{G}}=\mathcal{G}^*)$ to one, while weaker reconstructability or intervention-induced mean contrasts slow down recovery.

\section{Simulation Evaluations}\label{sec:simu}

\subsection{Setup}\label{subsec:simu-setup}

\paragraph{Data generation.} We consider the latent linear Gaussian structural model described in Section~\ref{sec:method} and examine the performance of different methods under different graph densities, intervention strengths, sample sizes, and intervention mechanisms.

The underlying causal mechanism $A$ was generated from an Erd\H{o}s--R\'enyi model under a random topological ordering to ensure acyclicity. 
For each ordered pair of nodes $i<j$, a directed edge $i\to j$ was added independently with probability such that the expected average node degree of $d$, where $d=2$ for the low-density setting and $d=4$ for the high-density setting. 
Nonzero edge weights were drawn uniformly from $\pm[0.3,0.6]$.

The control intercept was fixed at
$\eta^{(0)}=(-0.5,\ldots,-0.5)$. We consider two types of soft interventions---mean-shift and coefficient-shift interventions---as well as hard interventions, as described in Section~\ref{sec:method}, with matched interventio strength $\gamma_m\in\{-2,-4\}$. For coefficient-shift interventions, the coefficients in the intervention target's structural equation are reduced by one half, whereas for hard interventions they are set to zero. 

We sampled
$L_i\sim\mathrm{Lognormal}(\log 10,0.1^2)$ and
$C_i\sim\mathrm{Unif}(1,2)$, with
$s_j(C_i)=\exp(C_i\beta_j)$,
$\beta_j=-0.5g_j$, and
$g_j\sim\mathrm{Unif}(0.8,1.2)$. To introduce latent confounding, we consider pervasive low-rank confounders by setting the number of unmeasured factors to $n_{\mathrm{uc}}=2$ and generating a loading matrix $U\in\mathbb{R}^{p\times n_{\mathrm{uc}}}$ with i.i.d. entries from $\mathcal{N}(0,\sigma_{\mathrm{c}}^2)$, where $\sigma_{\mathrm{c}}$ controls the confounding strength. The noise covariance in all environments is specified as
$\Sigma_e^{(0)}=\operatorname{diag}(\operatorname{Unif}(0.5,0.8))+U U^{\top}$, where the low-rank term $UU^\top$ induces residual correlations across genes. We consider $\sigma_{\mathrm{c}}\in\{0,0.25,0.5\}$, graph dimension $p\in\{50,100\}$, control sample size $n_0=5000$, and interventional sample size $n_m\in\{100,200,500\}$ for each $m\in[p]$.

\paragraph{Competing methods.}
We compare PLNi with a range of existing methods for DAG learning. NOTEARS \citep{zheng2018dags}, DAGMA \citep{bello2022dagma}, LiNGAM \citep{Shimizu2006lingam}, MRS-PoissonSEM \citep{park2019high}, and Anchor-PLN-PC \citep{saeed2020anchored} learn causal structure from observational data. Among them, MRS-PoissonSEM is specifically designed for Poisson count data, whereas Anchor-PLN-PC accommodates Poisson--lognormal observations while accounting for measurement error and sample heterogeneity. DOTEARS \citep{xue2025dotears}, GIES \citep{hauser2012characterization}, IGSP \citep{wang2017permutation}, and GrIVET \citep{chen2024grivet} leverage interventional information. GrIVET is an IV-based local approach for linear models with latent confounding; we adapt its intervention-target identification step to our known single-target setting while retaining the subsequent estimation procedure.

For a fair comparison, library-size and covariate effects were adjusted for all methods using procedures compatible with their respective data models. For continuous-data methods, preprocessing included library-size normalization, a $\log(1+x)$ transformation, and adjustment for the observed covariate. Count-aware methods incorporated library size and covariates directly through their corresponding count or measurement models. Detailed preprocessing and implementation procedures are provided in Supplementary Section~A.1.

For methods returning continuous adjacency estimates, we use $\tau=0.1$. For MRS-PoissonSEM, which is more sensitive to thresholding, we use $\tau=0.02$ to provide a favorable evaluation of this comparator. Sensitivity to the evaluation threshold over a broad range of values is reported in Supplementary Section~B.1.2.

\paragraph{Evaluation metrics.}
We assess graph recovery using precision, recall, F1 score, and Structural Hamming Distance (SHD). For partially directed outputs from Anchor-PLN-PC and GIES, unresolved edges are oriented according to the true DAG only for scoring, favoring these methods. Let TP, FP, and FN denote the numbers of correctly identified, falsely identified, and missed directed edges, respectively. Then
$\mathrm{Precision}=\mathrm{TP}/(\mathrm{TP}+\mathrm{FP})$, $\mathrm{Recall}=\mathrm{TP}/(\mathrm{TP}+\mathrm{FN})$, and $\mathrm{F1}=2\,\mathrm{Precision}\times\mathrm{Recall}/(\mathrm{Precision}+\mathrm{Recall})=2\,\mathrm{TP}/(2\,\mathrm{TP}+\mathrm{FP}+\mathrm{FN})$.
SHD counts missing or extra adjacencies and incorrect orientations, with each reversal counted once. Higher F1 and lower SHD indicate better graph recovery.

\paragraph{Additional simulation studies.}
{ Section~B.1 provides complete results for the main simulation study, together with threshold sensitivity, comparisons across intervention mechanisms, and sensitivity to the log transformation used for continuous-data methods. Sections~B.2--B.4 examine heterogeneous perturbation effects, focused modeling ablations, and robustness to alternative count models, latent--measurement dependence, structural-model misspecification, and departures from Gaussianity. Sections~B.5--B.9 further investigate incomplete intervention coverage, numerical convergence and acyclicity, sensitivity to initialization and tuning, graph recovery in higher dimensions, and computational performance.}

\subsection{Results}\label{subsec:simu-results}
\begin{figure}[H]
    \centering
    \includegraphics[width=0.9\linewidth]{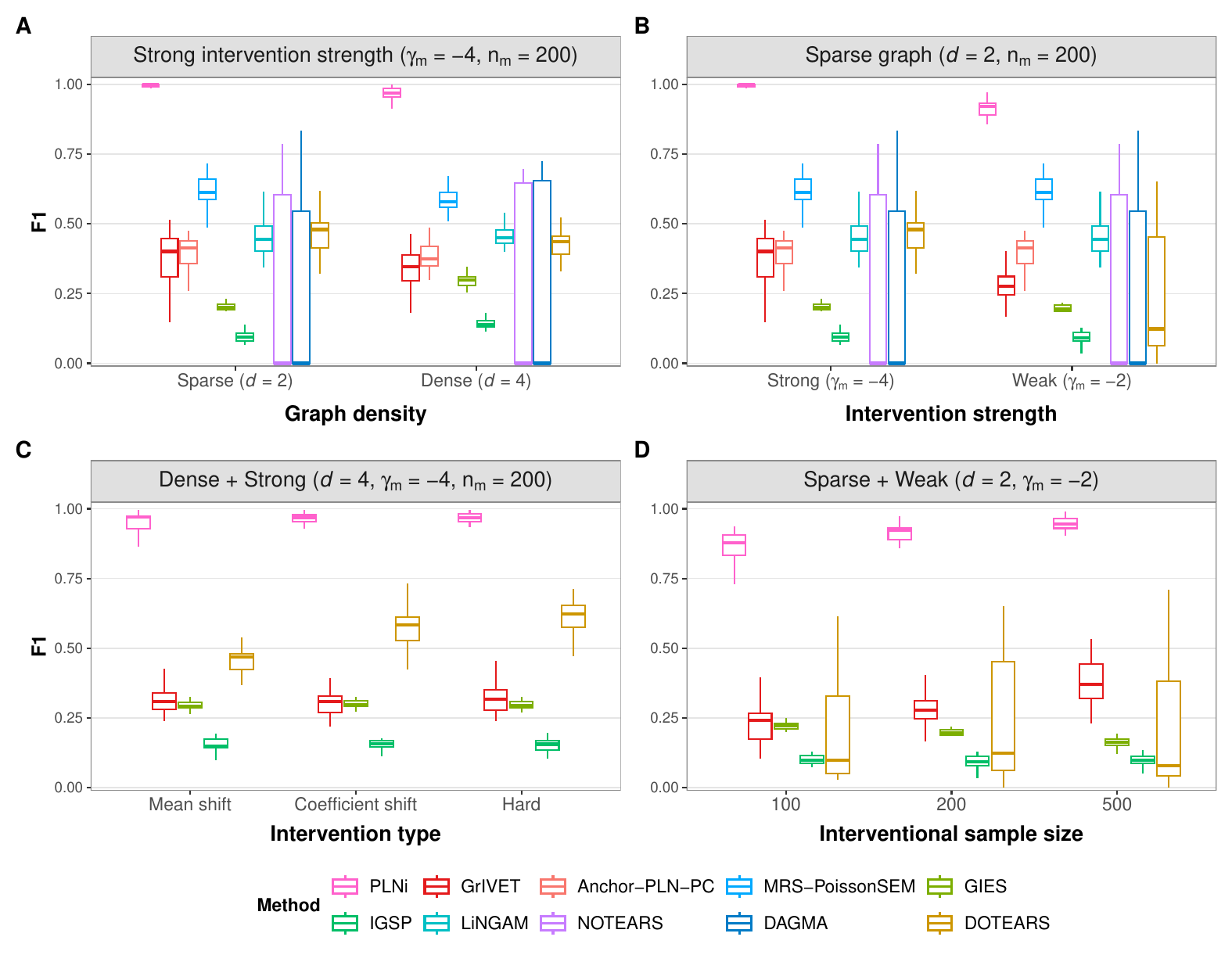}
    \caption{F1 score of different methods under different graph densities, different intervention strengths, intervention types, and varying sample sizes of interventional data.}
    \label{fig:F1}
\end{figure}

\begin{figure}[H]
    \centering
    \includegraphics[width=0.9\linewidth]{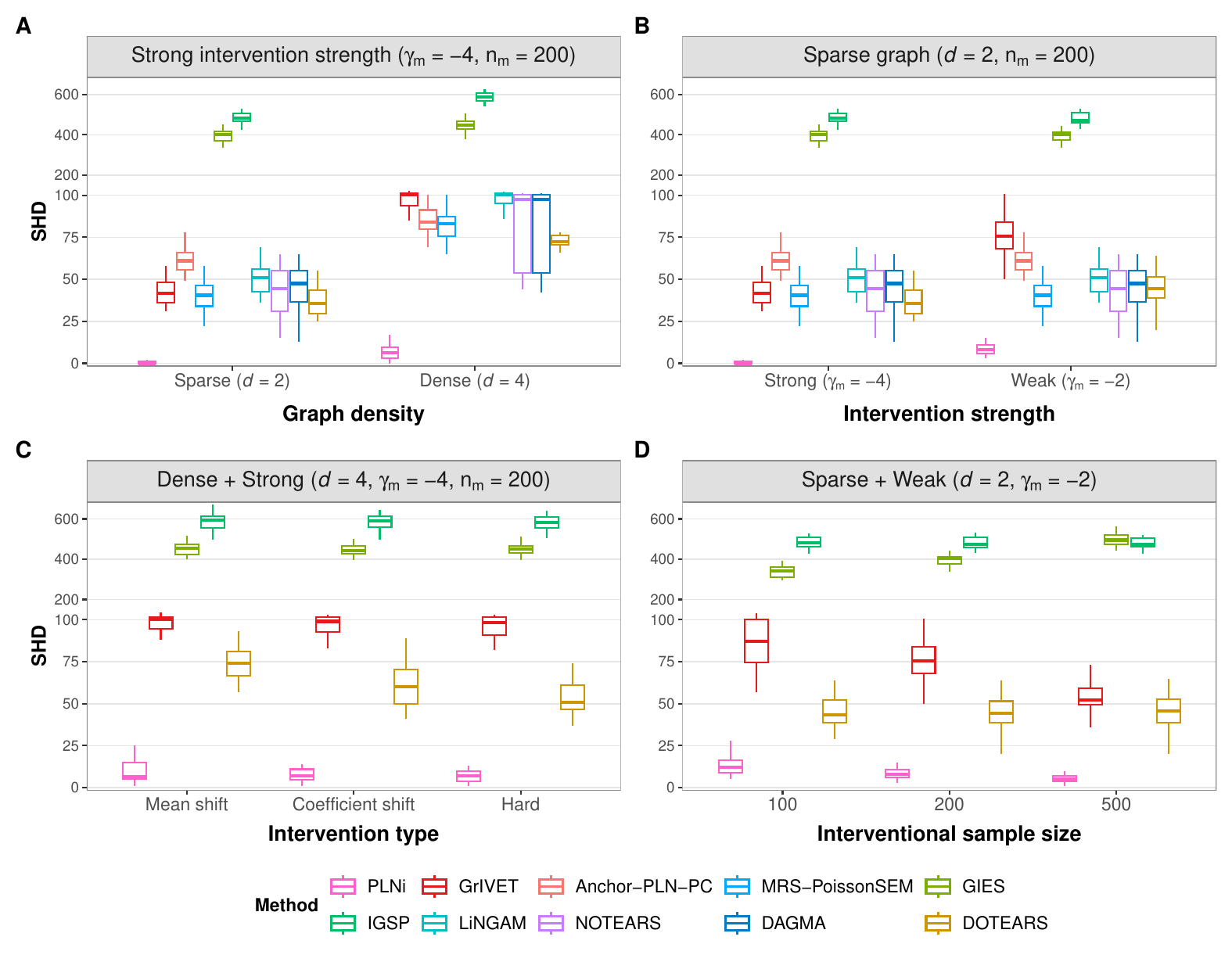}
    \caption{SHD of different methods under different graph densities, different intervention strengths, intervention types, and varying sample sizes of interventional data.}
    \label{fig:SHD}
\end{figure}

We report in the main paper the F1 score and SHD with graph dimension $p=50$ and $\sigma_{\mathrm{c}}=0.25$. 
Figures~\ref{fig:F1} and~\ref{fig:SHD} summarize the performance of different methods across varying graph densities, intervention strengths, intervention types,  and interventional sample sizes, measured by F1 score and SHD, respectively. We highlight several key observations below.

First, the performance of PLNi improves systematically with stronger interventions, lower graph density, and larger interventional sample sizes, as reflected by higher F1 scores and lower SHD. Its performance is also stable across intervention mechanisms, with slightly better recovery under hard interventions. These trends are consistent with  Theorem~\ref{thm:graph-recovery}.

Second, compared with observational methods, PLNi consistently achieves higher F1 and lower SHD across the considered settings. Among these methods, MRS-PoissonSEM generally performs best, followed by LiNGAM and Anchor-PLN-PC. Nevertheless, Anchor-PLN-PC, which uses the same Poisson--lognormal measurement model and measurement-error correction as PLNi, remains substantially less accurate than PLNi because it relies only on observational data. This comparison highlights the additional information provided by intervention-induced latent mean contrasts for identifying the directed causal graph.

Third, among the interventional methods, PLNi consistently achieves the best graph-recovery performance across the considered settings. GrIVET and DOTEARS generally perform better than the other interventional baselines, although their relative performance varies across settings. Both remain clearly below PLNi. Supplementary runtime comparisons further show that PLNi is substantially faster than GrIVET and is generally comparable to or faster than DOTEARS across the considered settings.

Lastly, several interventional-data-based methods, including DOTEARS, GIES, and IGSP, can even underperform the observational MRS-PoissonSEM. This may be attributed to two factors. First, some methods are better suited to hard or do-type interventions; accordingly, DOTEARS improves under hard interventions. Second, these methods do not explicitly model the count-valued measurement process, which may limit their performance relative to count-aware approaches such as MRS-PoissonSEM and Anchor-PLN-PC. These results further highlight the importance of appropriately modeling count-valued observations in causal discovery for perturbation studies.

\section{Application to Perturb-seq dataset}\label{sec:appl}

 We analyze a single-cell CRISPR perturbation dataset from the Perturb-seq study of \citet{replogle2022mapping}, focusing on the K562 chronic myelogenous leukemia cell line. In this experiment, K562 cells expressing a CRISPR interference (CRISPRi; dCas9–KRAB) system were subjected to pooled perturbations targeting approximately 2,000–2,500 essential genes. Single-cell RNA sequencing with guide capture was used to link each perturbation to its transcriptomic response, enabling systematic characterization of gene regulatory effects at single-cell resolution. After preprocessing to remove genes with insufficient perturbed-cell counts, extreme sparsity, or   {less than 50\% on-target knockdown}, 724 genes remain, of which we retain the 200 most variable genes based on log1p-transformed counts in control cells for our analysis. The resulting dataset contains 10,691 control cells and an average of 188 perturbed cells per intervention, with a minimum of 101 perturbed cells per gene. We use the gem-group as the observed technical batch covariate $C$.

To construct a biologically informed reference graph, we employed a transcription factor–target interaction network derived from ChIP-seq data \citep{chevalley2025large}, with interactions restricted to those relevant to or observed in the K562 cell line. Within the selected 200 genes, this reference network contains 504 directed edges, with a graph density of 0.0127. We compare PLNi with the competing approaches described in Section~\ref{sec:simu}. Detailed preprocessing and implementation are provided in Supplementary Material Section~C.1.

The ChIP-seq–derived network provides a useful but inherently limited reference for evaluating the learned DAG from perturb-seq data. While it offers independent evidence of regulatory relationships, it reflects physical TF–DNA binding rather than functional or causal effects on gene expression. Thus, agreement supports biological plausibility, particularly for direct interactions, whereas discrepancies may arise from context-specific regulation, indirect effects, or incomplete ChIP-seq coverage. Accordingly, this comparison should be viewed as a complementary, rather than definitive, validation of causal structure.

\subsection{Global recovery}\label{subsec:appl-PR-curve}
\begin{figure}[H]
    \centering
    \includegraphics[width=0.8\linewidth]{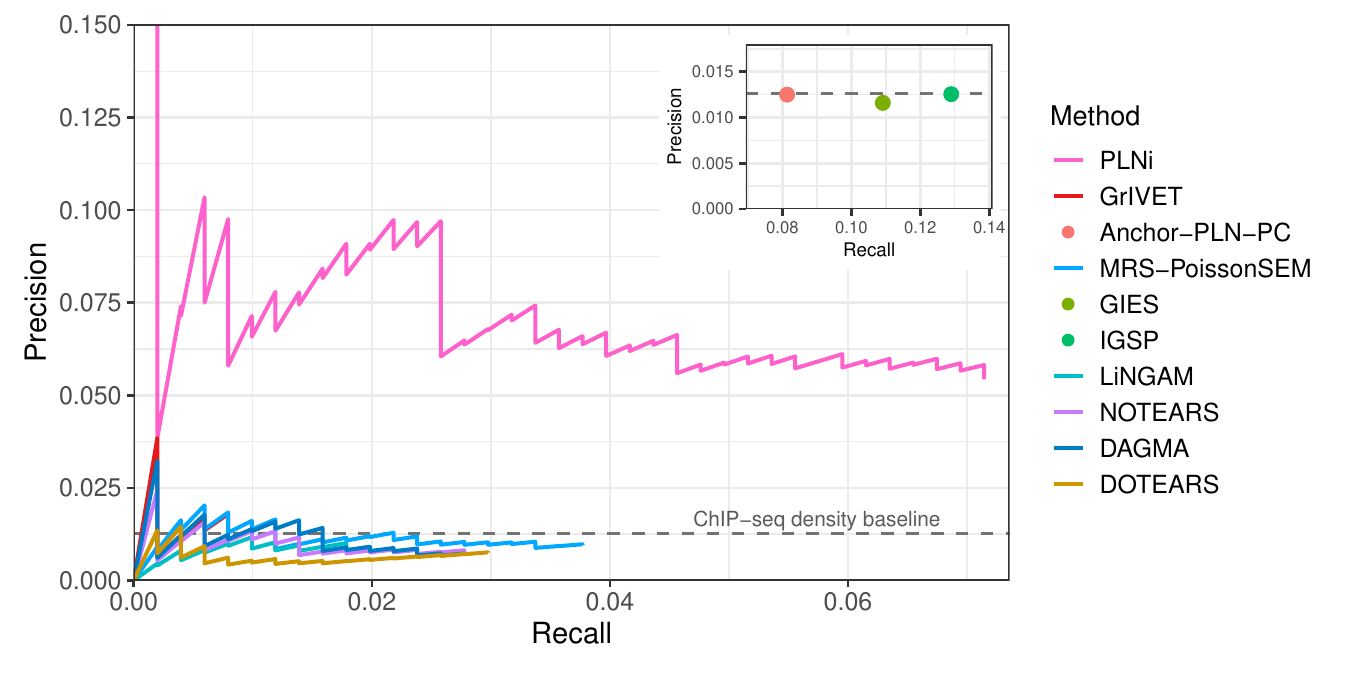}
    \caption{Ranked precision-recall comparison against the K562 ChIP-seq reference network.
}
    \label{fig:chipseq-pr-curve}
\end{figure}

{ Figure~\ref{fig:chipseq-pr-curve} presents ranked precision-recall curves for the different methods against the K562 ChIP-seq reference network.} For methods that return a weighted adjacency matrix, directed edges are ranked by the absolute values of their estimated coefficients, and precision and recall are computed by progressively including the highest-ranked edges, up to the top 2,000 nonzero edges. For methods that return only a binary graph (IGSP, GIES, and Anchor-PLN-PC), the estimated graph is represented by a single point. The dashed horizontal line indicates the precision expected from the density of the ChIP-seq reference network.

Figure~\ref{fig:chipseq-pr-curve} reveals two main findings. First, PLNi achieves substantially higher precision than the competing methods over a broad range of ranked edge sets. Its precision remains well above the ChIP-seq density baseline as additional edges are included, while recall increases steadily. In contrast, the competing weighted-graph methods generally attain substantially lower precision, with their curves remaining close to the density baseline. These results suggest that PLNi more effectively prioritizes regulatory relationships supported by independent ChIP-seq evidence.

Second, among the competing weighted-graph methods, DAGMA, NOTEARS, and GrIVET show modestly higher precision at small edge budgets, while MRS-PoissonSEM performs comparatively better as more edges are included. Among the binary-graph methods, IGSP achieves the highest recall, although its precision remains close to the density baseline. LiNGAM and DOTEARS generally exhibit lower precision and recall.

{ We further assessed network stability using 50 condition-stratified bootstrap resamples, comparing the interventional methods along with the global observational method LiNGAM. For all nonzero directed edges, the median pairwise Jaccard similarities were 0.253, 0.259, 0.375, 0.156, 0.667, and 0.128 for PLNi, GrIVET, GIES, IGSP, DOTEARS, and LiNGAM, respectively. Bootstrap-frequency-ranked ChIP-seq validation in Supplementary Section~C.2 also shows stronger enrichment for PLNi than for competing methods.}

\subsection{Distributional shifts in parent and child genes}\label{subsec:appl-KS-p}

{ To examine whether the inferred causal directions are supported by held-out interventional responses, we performed a five-fold held-out analysis. Cells within each perturbation and control condition were partitioned across five folds, with the causal graph estimated on four folds and distributional shifts evaluated on the held-out fold.} For methods returning weighted adjacency matrices, we constructed the evaluation graph using the 500 highest-ranked nonzero edges in absolute weight, retaining all edges when fewer than 500 were identified. Methods returning binary graphs were evaluated using their directed edges. For each perturbed gene, we identified its parent and child genes according to the estimated graph. We then used the two-sample Kolmogorov--Smirnov (KS) statistic to quantify expression distribution shifts between the corresponding perturbation and control conditions. Under a correctly inferred causal structure, child genes are expected to exhibit stronger distributional shifts, reflected by larger KS statistics.
\begin{figure}[H]
    \centering
    \includegraphics[width=0.8\linewidth]{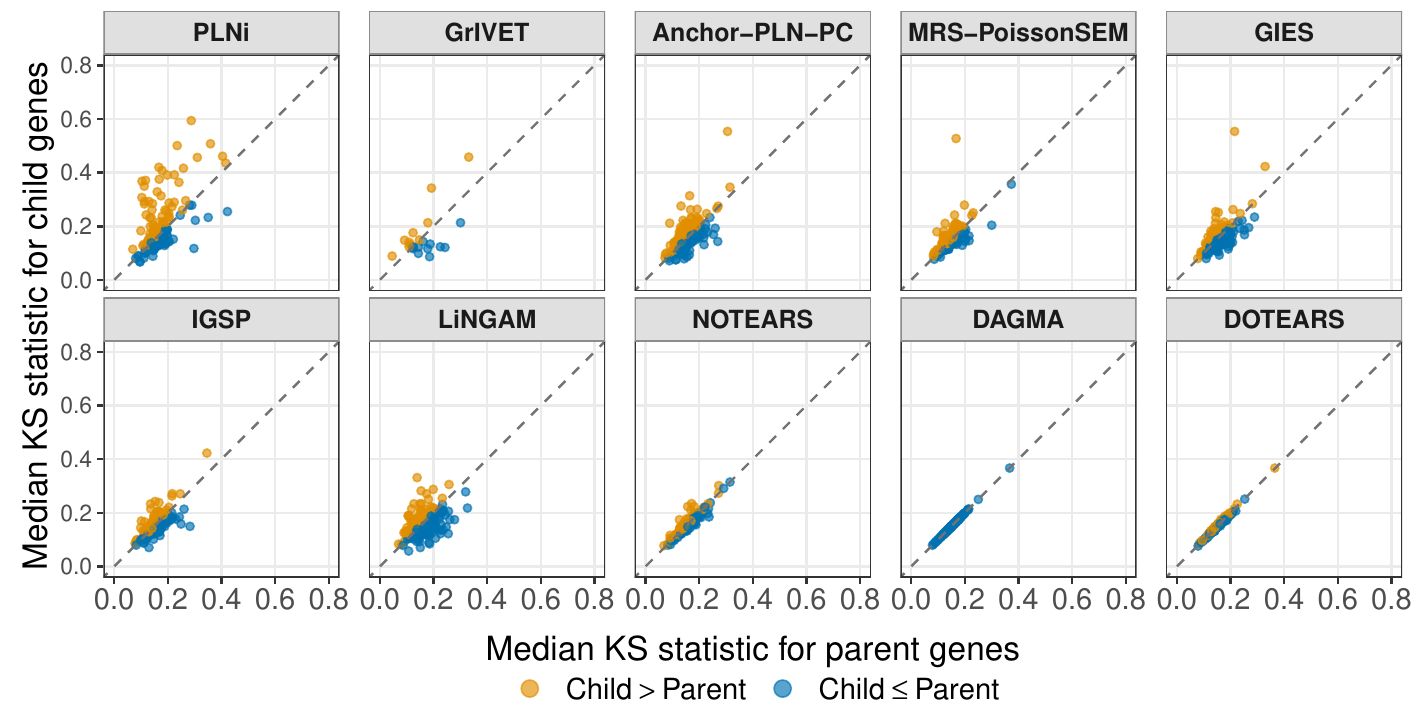}
    \caption{Held-out perturbation-level comparison of distributional shifts in inferred parent and child genes, shown for one held-out fold. For each perturbation, median KS statistics are computed over inferred parent and child genes. Each point represents one perturbation.
}
    \label{fig:perturbation_ks_D}
\end{figure}

As shown in Figure~\ref{fig:perturbation_ks_D}, for PLNi, most points lie above the diagonal line and toward the upper-left region of the panel, indicating that child genes generally exhibit stronger perturbation-induced distributional shifts than parent genes. In contrast, for the competing methods, many points cluster closer to the diagonal or near the origin, suggesting weaker or less distinguishable differences between parent and child gene sets. Cross-fold results in Supplementary Section C.3 confirm the strongest directional separation for PLNi, while a complementary edge-level analysis shows similarly strong directional evidence for PLNi and GrIVET, with PLNi exhibiting smaller cross-fold variation.

\subsection{Inferred gene regulatory network}\label{subsec:appl-bio-exp}

Inspection of the inferred network reveals several highly connected genes, including UBC, HSP90B1, SNRPA1, GATA1, and MYBL2. Among these, GATA1 is a well-established transcriptional regulator in hematopoiesis \citep{ferreira2005gata1}, and several of its outgoing edges are supported by ChIP-seq data, lending support to the biological plausibility of the inferred structure. MYBL2 also emerges as a source-like regulator, consistent with its established role in cell-cycle progression and proliferation \citep{musa2017mybl2}. In addition, HSPA5 has a strong inferred effect on HSP90B1. Previous perturbation experiments have shown that loss of HSPA5/BiP alters HSP90B1/GRP94 expression, providing independent support for a functional relationship between the two genes \citep{kharabimasouleh2014upr}. The inferred network visualization and more detailed biological interpretation are provided in Supplementary Section C.4.

\section{Discussion}\label{sec:disc}

This paper studies causal discovery from interventional count data motivated by single-cell perturbation experiments such as Perturb-seq, where discrete observations, measurement noise, and latent cellular heterogeneity pose key challenges. We propose a latent structural framework separating causal mechanisms on latent gene expression from the Poisson measurement process. Under one-sparse mechanism shifts, we establish identifiability through cross-environment latent mean contrasts and develop a moment-based estimator for sparse DAG recovery with finite-sample guarantees. Simulations and Perturb-seq analysis demonstrate the benefits of jointly accounting for count structure and interventional information.

Several extensions are possible. First, the identification principle may extend to broader settings in which latent mean contrasts remain identifiable, including alternative measurement models and less structured latent measurement models beyond the current Poisson--lognormal formulation. Second, higher-order contrasts may provide additional identifying information. Establishing analogous identifiability results and developing reliable estimation procedures for these broader settings remain interesting directions.

\section*{Data Availability}
The data that support the findings of this study are openly available at 
https://gwps.wi.mit.edu.

\section*{Acknowledgements} This research is supported by NIH grants GM129781 and HG013841.
\section*{Supplementary Materials} The supplementary material includes additional simulation details and evaluations,  detailed results for real data application and all the proofs. 

\bibliographystyle{chicago}

\bibliography{Bibliography-MM-MC}
\end{document}